\documentclass[prb,aps,twocolumn]{revtex4}
\usepackage{graphicx,color,latexsym}
\usepackage{dcolumn}
\usepackage{amsmath,amssymb,epsf,bm}
\begin{document}
\title{Interaction of  magnons with plasmons in a system of antiferromagnetic insulators coupled to a superconductor microwave cavity through the interfacial exchange interaction}
\author{A.~G. Mal'shukov}
\affiliation{Institute of Spectroscopy, Russian Academy of Sciences, Troitsk, Moscow, 108840, Russia}
\affiliation{Moscow Institute of Physics and Technology, Institutsky per.9, Dolgoprudny, 141700 Russia}
\affiliation{National Research University Higher School of Economics, Myasnitskaya str. 20, Moscow, 101000 Russia}
\begin{abstract}
A superconductor can interact with magnets through the magnetic field which is induced by the supercurrent. However, this interaction is rather weak and requires relatively large magnets to reach a sizable hybridization of  microwave electromagnetic modes with magnons. The size of the device can be drastically reduced by means of the large interfacial exchange interaction between spins in a magnet and a superconductor. In the spin-orbit coupled superconductor this interaction can be transmitted to the  s-wave condensate through a  mixing of triplet and singlet Cooper pairs.  In this paper we theoretically consider thin  antiferromagnetic films, which make a spin compensated contact to a superconducting wire, which forms a cavity for Mooij-Sch\"{o}n plasma waves. The effective Lagrangian is derived where  the  interaction of these waves with magnons is taken into account. Frequencies of hybrid magnon-plasmon modes are calculated. A system of two micromagnets placed far apart are also considered and their mutual coupling through the cavity mode is calculated.
\end{abstract}
\maketitle

\section{Introduction}

Hybrid systems incorporating superconducting and magnetic materials have received much attention recently. Such systems demonstrate unusual physical properties with a great potential for applications in quantum computing and information processing \cite{Qi,Linder}. A number of experimental and theoretical works were focused on cavity circuit QED systems where superconductor circuits couple to magnons in micromagnets \cite{Blanter,Hou,Haygood,Baity}, as well as to impurity spins \cite{Schuster,Kubo,Zhu}. This coupling usually takes place due  to the magnetic dipole interaction of  spins with the magnetic field which is induced by the supercurrent. However, this interaction is rather weak. Therefore, the sufficiently strong coupling can be achieved only for micromagnets sized up to tens and hundreds microns. On the other hand, at a good epitaxial contact, spins of electrons in the superconductor and magnets may interact with each other through the exchange interaction, which by many orders of magnitude exceeds the magnetic dipole coupling. Moreover, the spin-orbit coupling (SOC) of superconducting electrons helps to transmit the spin excitation to the s-wave condensate of singlet Cooper pairs. As was first shown by Edelstein \cite{Edelstein}, the SOC and the Zeeman splitting $\Delta_Z$, which is produced by the exchange interaction, result in a combined effect on the phase of the superconducting order parameter. This effect shows up in  the specific helix phase of the superconductor \cite{Samokhin,Barzykin,Agterberg,Kaur,Agterberg2,Dimitrova}, as well as in spontaneous supercurrents around ferromagnetic  insulator particles which are in a contact with the superconductor. \cite{Malsh island,Pershoguba,Hals}

These works are related to static thermodynamically equilibrium systems. In this situation, if spins of a micromagnet are ordered ferromagnetically on the interface with the superconductor, they may produce a too strong Zeeman effect on superconductor's electrons by suppressing the superconductivity when $\Delta_Z\gtrsim |\Delta|$, \cite{Chandrasekhar,Clogston} where $2|\Delta|$ is the superconducting gap. For very thin superconducting films this effect imposes a restriction on an admissible strength of  the  exchange field. On the other hand, its high strength is necessary to provide a good coupling between the superconducting condensate and magnetic excitations of the considered hybrid system. Alternatively, if spins on the interface are ordered antiferromagnetically, as in the case of an antiferromagnetic insulator having the compensated interface with the superconductor, the macroscopic Zeeman field is zero. Therefore, even a strong exchange interaction does not destroys the superconducting order parameter. While the static staggered spin magnetization does not produce a macroscopic Zeeman field, its dynamical part, which is associated with magnons, contributes to the time dependent $\Delta_Z$. This magnetization is weak and its influence on $|\Delta|$ can be neglected, while the effect on the phase of the order parameter should be taken into account, because it results in the coupling between a superconducting microwave cavity and an antiferromagnet.

The active part of the proposed set up, which is schematically shown in Fig.1, is represented by one or several bilayer films. Each of them is composed of  thin  antiferromagnetic insulator and superconductor films. These films, in turn, may consist of several atomic layers, down to a single layer. The thinnest possible films of both materials are desirable to enhance the interface effect of the exchange interaction. Moreover, the so called Rashba \cite{Rashba} SOC in many cases originates from interfaces.  For example, the strong SOC  and two-dimensional (2D) superconductivity was observed at some insulator's interfaces. \cite{Pavlov, Caviglia} A 2D Rashba-coupled superconducting film of Tl, Pb, Bi and In was grown on a semiconductor substrate  \cite{Yoshizawa,Matetskiy,Haviland,Zhang,Uchikashi,Sekihara}. Self-formation of a thin superconducting Rashba-coupled film at the interface of a Pb superconductor and a topological insulator was observed  \cite{Bai}, as well as some reports claim that the Dirac surface band migrates from the surface of a three-dimensional topological insulator towards the top of a thin superconducting over-layer and acquires a superconducting gap \cite{Sedlmayr,Trang}. In further calculations the width $W$ and the length $L$ of this rectangular bilayer film shall be taken within the submicron and micron ranges, respectively.

These bilayer systems may be built in a superconductor resonator. The latter can be a coplanar transmission line, as well as a nonlinear circuit which incorporates Josephson junctions, flux and charge qubits, etc.. Here, a simple model will be considered, where the superconducting resonator is represented by a thin and relatively long (10-20$\mu$m) closed wire, which supports Mooij-Sch\"{o}n \cite{Mooij} (MS) sound plasma  modes. The latter are controlled by the distributed capacitance and dynamic inductance of the wire. Its parameters  should be chosen appropriately, to bring the lowest resonance frequency $\omega_s$ well below the superconducting gap. The lowest magnon frequency $\omega_m$ is expected to be close to $\omega_s$. Since usually this frequency is rather high in antiferromagnets, the materials should be chosen appropriately. Namely, the superconductor must have large $\Delta$, while the antiferromagnetic  film must have the low N\'{e}el temperature and anisotropy energy.

The total system will be considered within a formalism of path integrals in the imaginary time. Such a formalism was previously used for studies of quantum fluctuations and phase slips  in thin superconducting wires. \cite{Schon,Zaikin1996,Golubev,Glazman} It is efficient for the calculation of the effective action by integrating out one-electron degrees of freedom. In our case such an effective action is expressed in terms of magnon variables and the superconductor's order parameter phase.

The article is organized in the following way. In Sec.II the effective action is derived for a superconducting wire which interacts with micromagnets through the exchange interaction. In Sec.III the Green function of the coupled magnon-cavity plasmon system is calculated for a single, as well as a pair of  antiferromagnets. The eigenfrequencies of these systems are calculated and the long-range interaction of micromagnets through the cavity plasmons is evaluated. The discussion of results is presented in Sec.IV.

\begin{figure}[tp]
\includegraphics[width=8.8 cm]{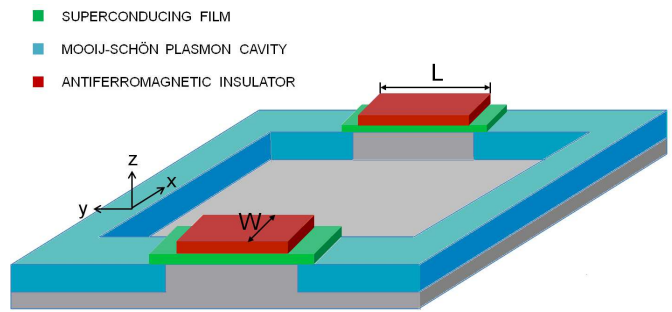}
\caption{(Color online) A schematic illustration of the system where the interaction of Mooij-Sch\"{o}n plasma waves with magnons in two antiferromagnetic insulators may be realized. The coupling of the supercurrent in the plasmon cavity to magnons occurs due to the interface exchange interaction of  spin-orbit coupled electrons in thin superconducting films with localized spins of antiferromagnets. The perimeter of the  rectangle is $L_{\mathrm{tot}}$. The coordinate  axes are locally oriented at each part of the wire, so that $x$ is directed along the wire} \label{fig1}
\end{figure}

\section{Effective action}

As it follows from previous studies, \cite{Malsh island,Pershoguba,Hals} in spin-orbit coupled superconductors a nonuniform static Zeeman field produces a spontaneous supercurrent. Therefore, in a certain sense a combined effect of SOC and the Zeeman splitting $\Delta_Z$ is  similar to the electromagnetic vector-potential, which couples to the phase $\theta$ of the superconducting order parameter. Our goal is to find an effective Lagrangian which is expressed in terms of the time dependent $\theta$ and the macroscopic dynamic spin polarization, by starting from the interface exchange interaction between superconductor's electrons and spins of the antiferromagnetic insulator. This can be done by averaging  the partition function of the whole system over electron degrees of freedom. A corresponding formalism, which is appropriate for superconducting circuits, was developed in Refs.[\onlinecite{Malsh island,Pershoguba,Hals}]. One should only modify it by taking into account SOC and the exchange interaction. The exchange interaction of $i$-th interface spin $\mathbf{S}_i$ with electrons is given by $J(\mathbf{r-r}_i)\bm{\sigma}\mathbf{S}_i$. By assuming that $J(\mathbf{r-r}_i)$ is peaked in the range of $|\mathbf{r-r}_i|$ smaller than the wavelength of electron's wave function, the one-particle exchange Hamiltonian can be written in the form
\begin{equation}\label{Hex}
H_{\mathrm{ex}}=J\sum_{i,\nu=1,2} \bm{\sigma}\mathbf{S}_{i\nu}\delta(\mathbf{r}-\mathbf{r_{i\nu}})\,,
\end{equation}
where $J=\int d^3r J(\mathbf{r-r}_i)$, the sum runs over two 2D sublattices $i1$ and $i2$ at the interface and $\bm{\sigma}$ is a vector composed of Pauli matrices $\sigma^j$  ($j=x,y,z$). In thermodynamic equilibrium all spins in each sublattice are equal to each other and oppositely directed in sublattices 1 and 2. Therefore, $\mathbf{S}_{i\nu}\equiv \mathbf{S}_{\nu}$ and $\mathbf{S}_{1}=-\mathbf{S}_{2}$. A simple model will be adopted of an uniaxial antiferromagnet whose easy axis is parallel to the unit vector $\mathbf{e}_z$, while the interface is located in the $x,y$ plane. Therefore, a localized spin may be represented in the form $\mathbf{S}_{i\nu}=S_{\nu}\mathbf{e}_z+\delta\mathbf{S}_{i\nu}$, where the second term describes dynamic and spatial variations  which are associated with magnons. By using the Holstein-Primakoff  transformation \cite{HolsteinPrimakoff} in the linear approximation, $\delta\mathbf{S}_{i\nu}$ can be expressed through bosonic operators. We further introduce the bosonic field $\delta\mathbf{S}(\mathbf{r})=\sum_{i\nu}\delta \mathbf{S}_{i\nu}\delta(\mathbf{r}-\mathbf{r_{i\nu}})$, so that the magnon's contribution to Eq.(\ref{Hex}) is written in the form $J\bm{\sigma}\delta\mathbf{S}(\mathbf{r})$.

Let us consider the partition function $Z$ of the hybrid bilayer system. The first step is to perform quantum mechanical and thermodynamic averaging of $Z$ over fast one-particle electron variables.  Further, one can express the rest as  a path integral of  $\exp(-\mathcal{A}_{\mathrm{eff}})$, \cite{Schon,Zaikin1996,Golubev} where $\mathcal{A}_{\mathrm{eff}}$ is the effective Euclidian action which depends on the collective variables: the order parameter $|\Delta|\exp i\theta$, the vector-potential $\mathbf{A}$, the scalar potential $V$, and the bosonic field of magnons. All these variables are functions of the imaginary time $\tau$ and spatial coordinates $\mathbf{r}$. The action $\mathcal{A}_{\mathrm{eff}}$ is given by
\begin{equation}\label{action}
\mathcal{A}_{\mathrm{eff}}=-\int_0^\beta d\tau\int d^n r \mathrm{Tr}(\ln G^{-1})+\mathcal{A}_\mathrm{m} + \mathcal{A}_{\mathrm{em}}\,,
\end{equation}
where $\beta=1/T$ is the inverse temperature (in units of energy), $d^n r$ denotes integration in the $n$-dimensional space, $\mathcal{A}_{\mathrm{m}}$ is the action associated with magnons, and $\mathcal{A}_{\mathrm{em}}$ is the electromagnetic part of the action which depends on $\mathbf{A}$ and $V$. $\mathcal{A}_{\mathrm{em}}$ provides a screening of the electric charge on the superconductor surface  by surrounding conductors. By this way it contributes to the distributed capacitance. A part of the action which depends on $|\Delta|$ is skipped, because $|\Delta|$ is weakly affected by magnons. The trace of $\ln G^{-1}$ is taken over spin and Nambu variables, while the function $ G^{-1}$ is given by
\begin{eqnarray}\label{G-1}
&&G^{-1}=\frac{\partial}{\partial\tau}+\tau_3\frac{\hat{k}^2}{2m}+\tau_3\alpha(\sigma^x \hat{k}_y-\sigma^y \hat{k}_x)-\tau_3\mu  + \nonumber\\
&&J\bm{\sigma}\delta\mathbf{S}(\mathbf{r},\tau)-i\tau_3eV(\mathbf{r},\tau)+e^{i\tau_3\theta(\mathbf{r},\tau)}\tau_1|\Delta|\,,
\end{eqnarray}
where $\mu$ is the  chemical potential, $\mathbf{\hat{k}}=-i\partial/\partial\mathbf{r}-(e/c)\tau_3\mathbf{A}(\mathbf{r},\tau)$ and the third term  represents the Rashba SOC. The Pauli matrices $\tau_i$, $i=1,2,3$, operate in the Nambu space.  It should be noted that the staggered static magnetization is not represented in Eqs.(\ref{action}) and (\ref{G-1}) because it gives only a weak correction to $|\Delta|$. For an evaluation of its effect let us consider a 2D superconductor, which is most subject to influence of interface spins. Let us assume an ideal Bragg reflection of electrons from the square lattice of such spins which is commensurate with the superconductor's lattice. The correction to $\Delta$ is given by the anomalous self-energy calculated in the second order on the exchange interaction. It is controlled by the small parameter  $(J/a^2)|E(\mathbf{k})-E(\mathbf{k+g})|^{-2}$, where $a$ is the 2D lattice period, $E(\mathbf{k})$ is the electron energy and  $\mathbf{g}$ is the Fermi surface nesting vector. This parameter will be assumed small, as long as the Fermi surface of the metal is sufficiently far from the nesting condition. In this case, by assuming $|E(\mathbf{k})-E(\mathbf{k+g})| \sim E_F$ and $ak_F \sim 1$, where $E_F$ and $k_F$ are, respectively, the Fermi energy and the Fermi wavevector, one can see that the staggered magnetization produces a weak effect on $\Delta$, as long as $JN_F\ll 1$, where $N_F$ is 2D state density of electrons at the Fermi surface. Note, that the same parameter $JN_F$ controls a position of  the Yu-Shiba-Rusinov localized state \cite{Yu,Shiba,Rusinov} inside the superconducting gap.

Since all four fields $\mathbf{A},V,\theta,\mathbf{S}$  are assumed small, it is possible to simplify considerably $\mathcal{A}_{\mathrm{eff}}$ by expanding $\ln G^{-1}$  up to the second order in these fields. Then, the partition function with the quadratic action may be integrated over $\mathbf{A}$ and $V$ in the semiclassical approximation. \cite{Schon,Zaikin1996,Golubev} The final result is expressed as the path integral over two independent variables: $\mathbf{S}(\mathbf{r},\tau)$ and $\theta(\mathbf{r},\tau)$. The calculation of the corresponding effective action $\mathcal{A}_{\mathrm{eff}}(\mathbf{S},\theta)$ is presented in Appendix A. Note, that the superconducting film is described here as a 2D metal, which might be a quantum well, or a monolayer. In this case Eq.(\ref{Hex}) should be averaged over $z$ with a corresponding wave function. Accordingly, $J$ should be redefined as $J \rightarrow J/d$, where $d$ is the effective thickness of the film.

 Besides the bilayer, the whole system includes also the plasma wave cavity which is represented by a thin and long closed wire. Therefore, the total action $\mathcal{A}_{\mathrm{tot}}(\mathbf{S},\theta)$ is a sum of actions, which are associated with the wire and one, or several hybrid bilayer films connected to the cavity. The total action has  the form
\begin{eqnarray}\label{Afin}
&&\mathcal{A}_{\mathrm{tot}}=-\int_0^\beta d\tau\int_0^{L_{\mathrm{tot}}} dx \left[\frac{\beta(x)}{2}\left(\frac{\partial\theta}{\partial\tau}\right)^2+\frac{\gamma(x)}{2}\left(\frac{\partial\theta}{\partial x}\right)^2\right. \nonumber \\
&&\left. +\int_0^W dy\left(\pi_{s\theta} \delta S_y(\mathbf{r},\tau)\frac{\partial\theta}{\partial x}+\right.\right.\nonumber\\
&&\left.\left.\int d^2 r^{\prime}\pi_{ss}(\mathbf{r-r}^{\prime}) \delta \mathbf{S}(\mathbf{r},\tau)\delta \mathbf{S}(\mathbf{r}^{\prime},\tau)\right)\right]+\mathcal{A}_\mathrm{m}\,,
\end{eqnarray}
where $L_{\mathrm{tot}}$ is the total length of the resonator  and bilayer films. Here, the first two terms represent the Lagrangian of MS plasma waves, \cite{Mooij} where the parameters $\beta(x)$ and $\gamma(x)$ take their respective values in the wire \cite{Schon,Zaikin1996,Golubev,Glazman} and in bilayers. These parameters determine the varying in space sound velocity $v_s=\sqrt{\gamma/\beta}$. The parameter $\beta=C/4e^2>0$, where $C$ is the capacitance per unit length. \cite{capacitance} In bilayer films $\gamma=Wn_s/4m$, where $n_s$ is the 2D density of superconducting electrons. In clean superconductors at low temperatures of interest the latter is close to the electron density. In the range of the resonator $\gamma=S n_{s\mathrm{3D}}/4m_{\mathrm{r}}$ where $S$ is the wire cross section, $n_{s\mathrm{3D}}$ is the three dimensional density of superconducting electrons in the wire and $m_{\mathrm{r}}$ is the electron mass there. The above relations for parameters remain valid also for dirty superconductors, by substituting corresponding expressions for $n_s$ in dirty systems \cite{Zaikin1996,Golubev}. Note, that the factor $\gamma$ represents the so called kinetic inductance of the resonator's wire. At the same time, the geometric inductance $L=2\ln (d/r_0)/c^2$, where $d$ is the cutoff and $r_0$ is the wire radius, is ignored. It can be neglected for thin wires if $\lambda_L^2\gg (S/2\pi)\ln (d/r_0)$, \cite{Zaikin1996,Golubev} where $\lambda_L$ is the London penetration depth.

The third term in Eq. (\ref{Afin}) represents the interaction between plasma waves and magnons. The corresponding coupling constant is $\pi_{s\theta}=JN_F\alpha n_s/n$. This constant is calculated in Appendix A  for a clean superconductor at the  low temperature $T\ll \Delta$. Smooth variations of the phase in time and space are assumed, whose characteristic frequencies and wavenumbers are much less then $\Delta$ and $\xi^{-1}$, respectively, where $\xi$ is the superconductor's correlation length. The SOC strength is taken sufficiently high, so that the corresponding splitting of electronic bands $\alpha k_F \gg \Delta$. At the same time, it is much less than the Fermi energy. All parameters were calculated in the leading order with respect to $\alpha/v_F$, where $v_F$ is the Fermi velocity. For a dirty superconductor  the parameter $\pi_{s\theta}$ can be taken from  Ref.[\onlinecite{Malsh vortex}], where the free energy of a spin-orbit coupled superconductor which interacts with a Zeeman field has been calculated. This free energy contains terms having a structure very similar to the third and fourth terms in Eq. (\ref{Afin}). Therefore, the parameters $\pi_{s\theta}$ and $\pi_{ss}$ can be extracted from this free energy. Here, we will focus on a simple case of a clean 2D superconductor.

Note, that since the wavelength of plasma waves is much larger than $L$ and $W$, the phase of the order parameter varies slowly in the range of a bilayer film. Consequently, only homogeneous magnon modes can efficiently interact with plasmons.

In the fourth term of Eq.(\ref{Afin}) one should take into account spatial nonlocality of the spin susceptibility $\chi_{ss}$, because fast varying staggered components of magnon modes might contribute to this term. By integrating over $\mathbf{r}$ and $\mathbf{r}^{\prime}$ it can be transformed to $\sum_{i\nu,j\mu} \delta\mathbf{S}_{i\nu}(\tau) \pi_{ss}(\mathbf{r}_{i\nu}-\mathbf{r}_{j\mu})\delta\mathbf{S}_{j\mu}(\tau)$.  As known, \cite{Anderson} in s-wave superconductors $\pi_{ss}(r)$ has a long-range part which exponentially decreases at distances  larger than the superconducting coherence length $\xi$. It also contains the oscillating short-range  Ruderman-Kittel-Kasuya-Yosida \cite{Ruderman,Kasuya,Yosida} interaction. At the same time,  antiferromagnetic magnons are combinations of staggered and smooth magnetizations. Since $\xi \gg a$, the staggered part can noticeably contribute to the spin-spin interaction only in the special case,  when a nesting of the Fermi surface at the wave vector of  the staggered magnetization takes place. In this case the spin susceptibility to the staggered field strongly increases. Otherwise, only the smooth magnetization $\mathbf{m}_i=(\delta\mathbf{S}_{i1}+\delta\mathbf{S}_{i2})/2$ should be taken into account. Within the adopted model of an easy axis antiferromagnet $\mathbf{m}$ is directed parallel to the $x,y$ plane. Since only long-wave magnons whose wavevectors are much larger than $\xi$ interact with cavity plasmons, one may transform the fourth term of Eq.(\ref{Afin}) to the local form: $\pi_{ss}\sum_{i} \mathbf{m}_i^2(\tau)$, where $\pi_{ss}=\sum_{j\mu}\pi_{ss}(\mathbf{r}_{i\nu}-\mathbf{r}_{j\mu})=-N_FJ^2(1-n_s/2n)/a^2$. As shown in Ref.[\onlinecite{GorkovRashba}], the latter is finite at $T\rightarrow 0$ due to SOC, which results in triplet Cooper pairs. They are polarized by the Zeeman field and in such a way contribute to the spin susceptibility. Note, that the latter  must vanish at $\alpha \rightarrow 0$ and $T=0$, while the above expression for $\pi_{ss}$ stays finite. It happens because it was calculated at $\alpha k_F \gg \Delta$.

The interaction $\pi_{ss}\sum_{i} \mathbf{m}_i^2$ could be added to the unperturbed antiferromagnetic action $\mathcal{A_{\mathrm{m}}}$. It is isotropic, as long as only parallel to the $x,y$ plane components of the magnetization are considered. In a general case one  may also take into account  the $m^z$ component. According to Ref.[\onlinecite{GorkovRashba}], the corresponding susceptibility $\pi_{ss}^z=-N_FJ^2/a^2$ is different from the parallel spin response $\pi_{ss}$. Hence, at the first sight it looks as if an important new term of the anisotropy energy appears in the magnetic energy of a very thin  antiferromagnetic film. However, the energy of anisotropy is determined mostly by the staggered part of the magnetization, which for typical antiferromagnets is much larger than the homogeneous magnetization associated with magnons. \cite{Akhiezer} Therefore, the additional anisotropy energy is by several orders of magnitude less than the initial energy and can be ignored.

\section{Hybrid plasmon-magnon resonances}

In this section hybridized modes of  MS waves and  magnons of an antiferromagnetic insulator shall be calculated for a single bilayer system attached to the MS resonator, as well as for two such systems at a distance from each other. In the latter case it is interesting to find out how strongly  these separated magnets may interact with each other through MS plasmons. Basically, as long as we are not interested in nonlinearities of the superconducting resonator and magnon system, the considered problem is reduced to a simple linear problem of coupled oscillators. Moreover, the system is far from a regime, which is characterized by strong  quantum fluctuations. For instance, an appropriate flux bias might be applied to the superconducting closed wire to tune up the system into a regime of the so called phase slip qubit. \cite{flux qubit} However, we face the linear problem, as long as instantons and quantum fluctuations are not important in the considered situation of a weakly perturbed superconducting wire. This problem is  more convenient to consider within the Hamiltonian formalism. By transforming the Lagrangian in Eq.(\ref{Afin}) to the real time representation one can write a corresponding Hamiltonian in the form
\begin{equation}\label{H}
H=H_{\mathrm{pl}}+H_{\mathrm{m}}+H_{\mathrm{int}},
\end{equation}
where the first two terms are unperturbed Hamiltonians of plasmons and magnons, respectively, while the third term corresponds to their interaction. $H_{\mathrm{pl}}$ is obtained from the Lagrangian given by the first two terms of  Eq.(\ref{Afin}), which represent, however, a system where parameters of plasma wave propagation vary in space. As a result, the eigenstates of the plasma oscillations are expressed by a superposition of the waves propagating in opposite directions. Therefore, the wave function of $n$-th mode has the form
\begin{equation}\label{thetan}
\theta_n=A_n(x)e^{ik_n(x) x}+B_n(x)e^{-ik_n(x) x}\,,
\end{equation}
where $A(x)$, $B(x)$  and $k_n(x)$ undergo a step-like change at boundaries between the wire and  bilayer films. Thus, at a given frequency $\omega_n$ the wavenumber varies in space according to the equation $k_n(x)=\omega_n/v_{s}(x)$. By transforming Eq.(\ref{Afin}) to the real time, the classical Hamiltonian of plasma excitations can be written  in the form
\begin{equation}\label{Hpl}
H_{\mathrm{pl}}=\int_0^{L_{\mathrm{tot}}} dx \left[\frac{P^2(x)}{2\beta} + \frac{\gamma}{2}\left(\frac{\partial\theta(x)}{\partial x}\right)^2\right] \,,
\end{equation}
where the momentum $P=\beta\dot{\theta}$ and the equation of motion is $\dot{P}=\partial_x(\gamma\partial_x\theta)$. From these expressions the wave equation for eigenfunctions of MS modes is obtained in the form
\begin{equation}\label{plwave}
\beta\omega_n^2\theta_n=\partial_x(\gamma\partial_x\theta_n).
\end{equation}
The boundary condition $\gamma_1\partial_x\theta_{1n}=\gamma_2\partial_x\theta_{2n}$ on the interface between the wire and the bilayer film follows directly from this equation, while the eigenfunctions $\theta_n(x)$ are continuous. From Eq.(\ref{plwave}) their orthogonality condition  can be written as
\begin{equation}\label{orthogonality}
\int_0^{L_{\mathrm{tot}}} dx\frac{\beta(x)}{\overline{\beta}}\theta_n(x)\theta^*_{n^{\prime}}(x)=\delta_{nn^{\prime}}\,,
\end{equation}
where $\overline{\beta}$ is the average of  $\beta$ over the wire length. It is added to the orthogonality condition to provide its standard form when $\beta(x)$ is constant in space.
In a quantized form  Eq.(\ref{Hpl}) can be written by representing $\theta$ and $P$ in terms of phonon destruction and creation bosonic operators, $b_n$ and $b^+_n$, respectively. Accordingly, we have
\begin{eqnarray}\label{quant}
\theta(x)&=&\sum_n \frac{1}{\sqrt{2\overline{\beta}\omega_n}}(\theta^*_n(x)b^+_n+\theta_n(x)b_n)\nonumber  \\
P(x)&=& i\sum_n \beta(x)\frac{\sqrt{\omega_n}}{\sqrt{2\overline{\beta}}}(\theta^*_n(x)b^+_n-\theta_n(x)b_n)\,.
\end{eqnarray}
It is easy to see that these functions satisfy the proper commutation relations $[P(x),\theta(x^{\prime})]=-i\delta(x-x^{\prime})$. From Eqs.(\ref{Hpl}),(\ref{orthogonality}) and (\ref{quant}), by ignoring the vacuum contribution, we finally obtain
\begin{equation}\label{Hfin}
H_{\mathrm{pl}}=\sum_n \omega_nb^+_nb_n\,.
\end{equation}
In the considered set up the length $L$ of bilayer films is much less than the length of the resonator's wire. Moreover, we are interested in the low-frequency regime, such that the wavelength of plasmons is much larger than $L$. In this case the presence of bilayer insertions produces only a weak effect on frequencies $\omega_n$ of the whole system. Therefore, by ignoring corrections of the order of $L/L_{\mathrm{tot}}$, the eigenfrequencies can be expressed as $\omega_n=\sqrt{\gamma_\mathrm{r}/\beta_\mathrm{r}}k_n$, where $k_n=2\pi n/L_{\mathrm{tot}}$ and the subscript $\mathrm{r}$ means that the parameters $\gamma$ and $\beta$ belong to the resonator. With the same accuracy one may simplify eigenfunctions Eq.(\ref{thetan}) of the system. Their exact wave numbers in the resonator's wire are given by $(2\pi n/L_{\mathrm{tot}}) + \delta k_n$, where $\delta k_n \sim L/L^2_{\mathrm{tot}}$ are small corrections. By ignoring these corrections it is possible to choose an appropriate linear combination of two functions, which correspond to a two-fold degenerate state, in such a way that they become two plane waves propagating in opposite directions, as it takes place in a closed wire without hybrid film insertions.

$H_{\mathrm{int}}$ can be obtained from the third term of  Eq.(\ref{Afin}). Since $\theta(x)$ varies slowly along the bilayer film, in the $i$-th film, by using the boundary condition, one may fix $\partial_x \theta(x)=(\gamma_r/\gamma_i)\partial_x \theta(x_i)$, where $x_i$ is taken just outside the  film. The factor $\gamma_r/\gamma_i$ takes into account the change of the phase derivative across the boundary between the resonator and the bilayer film. Further,  $\partial_x\theta(x_i)$ may be expanded  in eigenfunctions $\exp(ik_nx_i)/\sqrt{L_{\mathrm{tot}}}$ with $k_n=2\pi n/L_{\mathrm{tot}}$. In turn, for the slowly varying $\theta$ the spin density excitation $\delta\mathbf{S}(\mathbf{r})$ is dominated by the uniform in the $x,y$ plane magnon mode. Hence,   $\delta\mathbf{S}(\mathbf{r})$ is given by $\overline{\delta\mathbf{S}}$, where the overline denotes averaging over the interface. Note, that $\overline{\delta\mathbf{S}}$ still depends  on $z$. However, the lowest frequency mode is uniform in the $z$-direction. At the same time, due to confinement in thin films this mode is well separated from higher frequency nonuniform modes. Therefore, later only the uniform magnon will be taken into account. From all this, by taking into account Eq.(\ref{Afin}) and Eq.(\ref{quant}), we arrive to the interaction Hamiltonian in the form
\begin{equation}\label{Hint}
H_{\mathrm{int}}=\pi_{s\theta} WL\overline{\delta S_{y}}\sum_{i,n}\frac{\gamma_r}{\gamma_i}\frac{ik_n(e^{ik_nx_i}b_n-e^{-ik_nx_i}b^+_n)}{\sqrt{2L_{\mathrm{tot}}\overline{\beta}\omega_n}}
\end{equation}

\subsection{A single antiferromagnet coupled to the superconducting cavity}

 Let us consider a single bilayer film which is connected to the cavity at $x=0$. Accordingly, we set in Eq.(\ref{Hint}) $x_i=0$ and $i=1$. By defining the operator $\Lambda_n=ik_n(b_n-b^+_n)/\sqrt{2\overline{\beta}\omega_n}$,  Eq.(\ref{Afin}) can be written as
\begin{equation}\label{Hint2}
H_{\mathrm{int}}=g_{s\theta} \overline{\delta S_{y}}\sum_{n} \Lambda_n\,,
\end{equation}
where the coupling constant $g_{s\theta}=\pi_{s\theta}\gamma_rWL/\gamma_1\sqrt{ L_{\mathrm{tot}}} $. Let us define the   Green function of plasmons as
\begin{equation}\label{G}
G_{nn^{\prime}}(t-t^{\prime})=-i\langle \mathrm{T}[\Lambda_n(t)\Lambda_n(t^{\prime})]\rangle\,,
\end{equation}
where the average is taken over the ground state and $\mathrm{T}$ is the chronological operator. In the Fourier representation the unperturbed function $G_{0nn}$ can be obtained from the Hamiltonian Eq.(\ref{Hfin}) in the form
\begin{equation}\label{G0}
G_{0nn}(\omega)=-\frac{\beta_{\mathrm{r}}}{\gamma_{\mathrm{r}}\overline{\beta}}\frac{\omega^2_n}{\omega^2-\omega^2_n+i\delta}\,.
\end{equation}
In turn, the  Green function of magnons can be defined as
\begin{equation}\label{Gm}
\mathcal{G}_{\mathrm{m}}(t-t^{\prime})=-i\langle \mathrm{T}[\overline{\delta S_{y}}(t)\overline{\delta S_{y}}(t^{\prime})]\rangle\,.
\end{equation}
This function involves only spins at the interface with the superconductor. At the same time, it is possible to express $\mathcal{G}_{\mathrm{m}}$  in terms of the bulk Green's function $\mathcal{G}^{3D}_{\mathrm{m}}$ of the film, by employing the fact that only lowest-frequency uniform in the $z$-direction magnonic modes should be taken into account. Therefore, $\mathcal{G}_{\mathrm{m}}=\mathcal{G}^{3D}_{\mathrm{m}}$, where $\mathcal{G}^{3D}_{\mathrm{m}}$ is given by Eq.(\ref{Gm}), with $\overline{\delta S_{y}}$ substituted for $\sum_{i_z}\overline{\delta S_{y}}_{i_z}/N_z$. The sum runs over $N_z$ atomic layers of the film. The unperturbed bulk function $\mathcal{G}^{3D}_{0\mathrm{m}}$ can be obtained from the average spin density  $m_y(t)$ which is induced in the antiferromagnetic insulator  by a time dependent magnetic field $B_y(t)$, at $H_{\mathrm{int}}=0$. In the linear response approximation this density is given by
\begin{equation}\label{mt}
m_y(t)=N^2_z\frac{LW}{d_z}\int dt^{\prime}\mathcal{G}^{3D(R)}_{0\mathrm{m}}(t-t^{\prime})g\mu_BB_y(t^{\prime})\,,
\end{equation}
where the superscript  $R$ denotes the retarded function and $d_z$ is the thickness of the film. By expressing $m_y(t)$ through the magnetic susceptibility $\chi_{yy}$, as $m_y(t)=\chi_{yy} B_y/g\mu_B$ we obtain from Eq.(\ref{mt})
\begin{equation}\label{G02}
\mathcal{G}^R_{0\mathrm{m}}(\omega)=\frac{\chi_{yy}}{(g\mu_B)^2}\frac{d_z}{LWN_z^2}\,.
\end{equation}
In turn, the time ordered function Eq.(\ref{Gm}) at $\omega>0$ coincides with the retarded function. Therefore, Eq.(\ref{G02}) is also valid for it. The susceptibility $\chi_{yy}$ can be taken from literature. Let us consider an antiferromagnet with the uniaxial anisotropy, whose anisotropy axis is parallel to the $z$-axis. In this case, according to \cite{Akhiezer},
\begin{equation}\label{chi}
\chi_{yy}=\chi_0\frac{\omega^2_{\mathrm{m}}}{\omega^2_{\mathrm{m}}-\omega^2}\,,
\end{equation}
where $\omega_{\mathrm{m}}$ is the frequency of the doubly degenerate uniform magnon mode, and $\chi_0 \sim (g\mu_B)^2/ T_Na^3$, with $T_N$ denoting the Neel temperature. The Dyson equation for the function Eqs.(\ref{Gm}) follows from Eqs.(\ref{Hint2}) and (\ref{G}). It is written as
\begin{equation}\label{Dyson}
\mathcal{G}_{\mathrm{m}}=\mathcal{G}_{0\mathrm{m}}+\mathcal{G}_{0\mathrm{m}}g_{s\theta}^2\sum_nG_{0nn}\mathcal{G}_{\mathrm{m}}\,.
\end{equation}
Note, that the sum  runs over the two-fold degenerate states $n$ and $-n$. By substituting in this equation the coupling constant $g_{s\theta}$, from Eqs.(\ref{G0}), (\ref{G02}) and (\ref{chi}) we obtain the following equation for the eigenfrequency, which is given by the pole of $\mathcal{G}_{\mathrm{m}}$:
\begin{equation}\label{pole}
\omega^2_{\mathrm{m}}-\omega^2=\omega^2_{\mathrm{m}}\Gamma \sum_{n>0}\frac{\omega^2_n}{\omega^2_n-\omega^2}\,,
\end{equation}
where
\begin{equation}\label{Gamma}
\Gamma=2\frac{\chi_0}{ (g\mu_B)^2}(JN_F)^2\alpha^2\frac{aWL}{N_zL_{\mathrm{tot}}}\frac{\gamma_{\mathrm{r}}\beta_{\mathrm{r}}}{\gamma_{1}^2\overline{\beta}}\,.
\end{equation}
It was taken into account that  in clean superconducting films  $n_s \approx n$ at low temperatures.

If  resonator and magnon frequencies are close to each other, one can leave in the right hand side of Eq.(\ref{pole}) only a single term in the sum over $n$. If it is a term with $n=1$, then we must assume that $\omega_2-\omega_1=2\pi v_s/L_{\mathrm{tot}}\gg |\omega_1-\omega_{\mathrm{m}}|$, where the sound velocity $v_s=\sqrt{\gamma_{\mathrm{r}}/\beta_{\mathrm{r}}}$. In this case the frequencies of hybrid resonances are obtained from Eq.(\ref{pole}) in the form
\begin{equation}\label{omegamp}
\omega_{\mathrm{mp}}=\frac{\omega_{\mathrm{m}}+\omega_{1}}{2}\pm \frac{1}{2}\sqrt{(\omega_{\mathrm{m}}-\omega_{1})^2+\omega^2_{\mathrm{m}}\Gamma}\,.
\end{equation}
Therefore, the maximum splitting between two resonance modes reaches $\omega_{\mathrm{m}}\sqrt{\Gamma}$, which is a measure of the plasmon-magnon interaction strength. For its evaluation let us assume in Eq.(\ref{Gamma}) that $L=2\mu$m, $L_{\mathrm{tot}}=20\mu$m, $W=200$nm, $JN_F=0.1$, $\alpha/v_F=0.1$, $N_z=10$. Further, the ratio $\beta_{\mathrm{r}}/\overline{\beta}\approx 1$, because $\overline{\beta}\approx \beta_{\mathrm{r}}$ at $L\ll L_{\mathrm{tot}}$. The ratio $\gamma_{\mathrm{r}}/\gamma_{1}=S n_{s\mathrm{3D}}m/W n_{s}m_{\mathrm{r}}$. It may vary widely, depending on material parameters. Let us take the width of the wire the same as that of the bilayer, its thickness $t_{\mathrm{r}}=$10nm and $m=m_{\mathrm{r}}$. Then, $\gamma_{\mathrm{r}}/\gamma_{1}=n_{s\mathrm{3D}}t_{\mathrm{r}}/n_s$. If the resonator's wire is made of a dirty superconductor, we have $n_{s\mathrm{3D}}=\pi n_{\mathrm{3D}}\Delta_{\mathrm{r}}\tau_{\mathrm{imp}}$, where $\Delta_{\mathrm{r}}$ and $\tau_{\mathrm{imp}}$ are the order parameter and the mean electron scattering time in the wire, respectively. \cite{Kopnin} Hence, we obtain $\gamma_{\mathrm{r}}/\gamma_{1}=(n_{\mathrm{3D}}t_{\mathrm{r}}/n_s)(\pi\Delta_{\mathrm{r}}\tau_{\mathrm{imp}})$. Therefore, although the first factor in brackets is always large, the second one may be very small, so that the considered ratio may be of order 1. In the case when the superconductor in the bilayer film is represented by a proximized narrow gap doped semiconductor, such as InAs quantum well, one should take into account that the effective mass of electrons there is much less then in the metal wire. This results in decreasing of $\gamma_{\mathrm{r}}/\gamma_{1}$. At the same time, the electron density in the semiconductor by many orders of magnitude is less than in the metal, which leads to increasing of this ratio. So that there are many competing factors which compensate each other. By setting $\gamma_{\mathrm{r}}/\gamma_{1} \sim 1$, with chosen above parameters  we arrive to the evaluation $\Gamma = (E_F/T_N)n_sa^3\times 1.6\times 10^{-5}$. At $E_F/T_N=100$ and $a^3n_s =1$ the splitting of resonances is $\omega_{\mathrm{m}}\sqrt{\Gamma}=4\omega_{\mathrm{m}}\times 10^{-2}$.

It is expected that the dissipation in the resonator should be weak.  This imposes a restriction on $\omega_1$, such that $\omega_1=v_{s}(2\pi/L_{\mathrm{tot}})<2\Delta$. This means that  $L_{\mathrm{tot}}$ must be large enough. For a dirty  wire with the cross-section 10$^{-15}$m$^2$ and   $2\Delta\approx$ 0.5 meV the minimum permissible length $L_{\mathrm{tot}}$ was calculated \cite{Mooij} as large as 30$\mu$m, and  $v_s=3.3\cdot 10^{6}$ m/sec.

\subsection{Interaction of magnons via the cavity plasma waves}

In this subsection we consider two antiferromagnets  at the distance $L_{12}$ apart each other. In this case in Eq.(\ref{Hint}) the sum runs over  $i=1,2$. Let us assume that these magnets are placed at $x_1=0$ and  $x_2=L_{12}$, respectively. The Hamiltonian in Eq.(\ref{Hint2}) should be  modified accordingly, as
\begin{equation}\label{Hint3}
H_{\mathrm{int}}=g_{s\theta 1} \overline{\delta S_{1y}}\sum_{n} \Lambda_{1n}+g_{s\theta 2} \overline{\delta S_{2y}}\sum_{n} \Lambda_{2n}\,,
\end{equation}
where the coupling constants $g_{s\theta 1}$ and $g_{s\theta 2}$ are given by the same expressions as $g_{s\theta}$ in Eq.(\ref{Hint2}), where parameters $\pi_{s\theta},W, L$ and $\gamma$  are substituted by their respective values for corresponding bilayer films. The operators $\Lambda$ are given by $\Lambda_{1n}=ik_n(b_n-b^+_n)/\sqrt{2\overline{\beta}\omega_n}$ and $\Lambda_{2n}=ik_n[b_n\exp(ik_nL_{12})-b^+_n\exp(-ik_nL_{12})]/\sqrt{2\overline{\beta}\omega_n}$.  Plasmon's Green functions are defined as
\begin{equation}\label{Gij}
G^{ij}_{nn^{\prime}}(t-t^{\prime})=-i\langle \mathrm{T}[\Lambda_{in}(t)\Lambda_{jn^{\prime}}(t^{\prime})]\rangle\,,
\end{equation}
where $i$ and $j$ take on the values $1$ and $2$. The unperturbed functions are diagonal on $n$ and $n^{\prime}$. It follows from definitions of $\Lambda_{1n}$ and $\Lambda_{2n}$ that $G^{11}_{0nn}(\omega)$ and $G^{22}_{0nn}(\omega)$ are given by Eq.(\ref{G0}). At the same time,
\begin{equation}\label{G012}
G^{12}_{0nn}(\omega)=G^{*21}_{0nn}(\omega)=\frac{\beta_{\mathrm{r}}\omega_n}{\gamma_{\mathrm{r}}\overline{\beta}}\frac{\omega_n\cos k_nL_{12}-i\omega\sin k_nL_{12}}{\omega^2_n-\omega^2}\,.
\end{equation}

The Green's functions of magnons are given by
\begin{equation}\label{Gmij}
\mathcal{G}^{ij}_{\mathrm{m}}(t-t^{\prime})=-i\langle \mathrm{T}[\overline{\delta S^i_{y}}(t)\overline{\delta S^j_{y}}(t^{\prime})]\rangle\,.
\end{equation}
The unperturbed functions $G^{0ij}_{\mathrm{m}}$ are diagonal with respect to $i$ and $j$. Similar to the previous subsection they can be expressed in terms of  magnetic susceptibilities of both antiferromagnets. They are represented by Eqs. (\ref{G02}) and  (\ref{chi}), where parameters $L,W,N_z, \chi_0^{yy}$, and $\omega_{\mathbf{m}}$ are given by their  values in  magnets 1 and 2, respectively. The frequencies of antiferromagnets 1 and 2, which are made of the same materials, are equal, if the long-range interaction of magnetic dipoles is ignored. Otherwise, these frequencies depend on  shapes of  magnetic films due to their different depolarization factors. However, in antiferromagnets the corresponding corrections to magnon energies are usually small \cite{Akhiezer}. Nevertheless, they can be ignored only if they are much less than detuning between plasmons and magnons.

From  Eqs. (\ref{Hint3}-\ref{Gmij}) one can write the Dyson equation for $\mathcal{G}^{ij}_{\mathrm{m}}$ in the form
\begin{equation}\label{Dyson2}
\mathcal{G}^{ij}_{\mathrm{m}}=\mathcal{G}^{ii}_{0\mathrm{m}}\delta^{ij}+\mathcal{G}^{ii}_{0\mathrm{m}}g_{s\theta i} \sum_{l=1,2} D^{il} g_{s\theta l}\mathcal{G}^{lj}_{\mathrm{m}}\,,
\end{equation}
where $D^{il}=\sum_n G^{il}_{0nn}$. In the case when one of the plasmon frequencies, for instance the frequency of its lowest-frequency  mode $\omega_1$,  is close to the magnon frequencies $\omega_{\mathrm{m}1}$ and $\omega_{\mathrm{m}2}$, one may leave in $\sum_n G^{il}_{0nn}$ only the term with $n=1$. Then, from Eq.(\ref{Dyson2}) the equation for eigenfrequencies can be written  as
\begin{eqnarray}\label{det}
\left[\left(\omega^2_{\mathrm{m}1}- \omega^2\right)\left(\omega^2_{1}- \omega^2\right)-\omega^2_{\mathrm{m}1}\omega^2_{1}\Gamma_1\right]\times&&\nonumber\\
\left[\left(\omega^2_{\mathrm{m}2}- \omega^2\right)\left(\omega^2_{1}- \omega^2\right)-\omega^2_{\mathrm{m}2}\omega^2_{1}\Gamma_2\right]=&&
\nonumber\\\omega^2_{\mathrm{m}1}\omega^2_{\mathrm{m}2}\omega^4_{1}\Gamma_1\Gamma_2\left[1+\omega_1^{-2}(\omega^2-\omega_1^2)\sin^2 k_1L_{12}\right]\,,&&
\end{eqnarray}
where the coupling parameters $\Gamma_1$ and $\Gamma_2$ are defined in a similar way as in Eq.(\ref{pole}). The right-hand side of this equation determines the interaction between magnons. When it vanishes,  the solution of  Eq.(\ref{det}) is given by independent magnon-plasmon modes which are related to two antiferromagnets (see Eq.(\ref{pole})). The second term in brackets depends on the distance between them.  It is much smaller than the first term because $(\omega^2-\omega_1^2)/\omega_1^{2}\ll 1$. Therefore, it can be treated perturbatively. Let us consider Eq.(\ref{det}) in the special case when $\omega_{\mathrm{m}2}=\omega_{\mathrm{m}1}=\omega_1$. Then, the frequencies of four magnon-plasmon eigenmodes are given by
\begin{equation}\label{modes}
\omega_{\mathrm{mp}}=\omega_1\pm\frac{\omega_1}{2}\left[(\Gamma_1+\Gamma_2)\pm \frac{\Gamma_1\Gamma_2}{\sqrt{\Gamma_1+\Gamma_2}}\sin^2 \frac{2\pi L_{12}}{L_{\mathrm{tot}}}\right]^{1/2}\,.
\end{equation}
The first term in brackets does not depend on the distance between antiferromagnets. It corresponds to their independent spin dynamics and scales as $\sqrt{N}$, where N is the number of magnets in the system. It is well known behavior, which is typical for spin systems interacting with a cavity mode. The second term results in a splitting of the resonance, which appeared to be of higher order with respect to $\Gamma$. It varies with $L_{12}$ due to interference of plasma waves which scatter on spin excitations of two magnets. The plus and minus signs in the second term of Eq.(\ref{modes}) correspond to inphase and counterphase spin dynamics of antiferromagnets, respectively. Therefore, if $\Gamma_1 \sim \Gamma_2 \sim \Gamma$, the difference $\sim \Gamma^{3/2} \omega_1 \sin^2 (2\pi L_{12}/L_{\mathrm{tot}})$ in energies of these modes gives the  magnon's binding energy. This energy oscillates with the distance $L_{12}$. In the considered regime the oscillation amplitude does not depend on $L_{12}$.  It is a consequence of the one-dimensional propagation of the order parameter phase. For example, in a two-dimensional superconductor the condensate mediated spin-spin (static) interaction decreases as $L_{12}^{-2}$.\cite{Malsh spin-spin} However, one should take into account the exponential attenuation of plasma modes, which is caused by quasiparitcles and  inelastic scattering of electrons. At the same time, at low temperatures $k_BT \ll \Delta$ and at weak  inelastic scattering of electrons one may neglect this damping at small enough $L_{12}$.\cite{Mooij} At least, in the considered situation $L_{12}$ is less than the wavelength of the lowest energy plasmon.

\section{discussion}

Interaction of electrons in a superconducting microwave cavity with magnons of an antiferromagnetic insulator  has been studied. In contrast to the well known mechanism, which relies on the magnetic interaction between superconducting currents and magnetic moments of spins, here a basically different mechanism is suggested. It employs the combined effect of the spin-orbit coupling and the exchange interaction between itinerant electronic spins of a superconductor and localized  spins of an antiferromagnet. It results in the effective coupling between the supercurrent and magnons, which formally resembles the previously considered one. A crucial distinction, however, is that the former  is much stronger due to the large exchange interaction. For instance, this interaction causes a splitting of the cavity resonance frequency $\sim 4\cdot 10^{-2}\omega_{\mathrm{m}}$ which is produced by 200 nm$\times$2 $\mu$m magnetic film consisting of 10 atomic layers. This splitting increases for tinner films, because only spins at the interface are involved in the exchange interaction with superconductor's electrons. For this reason the proposed mechanism is most efficient for studying thin antiferromagnetic films and atomic monolayers, that can not be achieved by the conventional method, which involves much larger magnetic volumes. With the above used parameters, for an antiferromagnetic monolayer, the single-spin coupling energy $\omega_{\mathrm{spin}}$ can be evaluated as $\sim 10^{-1}\omega_{\mathrm{m}}/\sqrt{N}$, where $N$ is the number of spins in the monolayer. By assuming the distance between spins $\sim 5${\AA}, for  a 200 nm$\times$2 $\mu$m film, at $\omega_m=100$ GHz we obtain $\omega_{\mathrm{spin}}\sim 10$ MHz. It is important that such a strong interaction can be realized for antiferromagnets, whose magnetic susceptibility is usually quite low. Therefore, the proposed interaction mechanism can be employed  in the emerging field of antiferromagnetic spintronics \cite{Baltz}.

However, a practical realization of the discussed set-up is a challenge for experimentalists. The major problem is that frequencies of Mooij-Sch\"{o}n modes are rather high in wires of a reasonable length. Also, high is the magnon frequency in most of antiferromagnets. On the other hand, these frequencies must be smaller than the superconducting gap. Therefore, one must use superconductors with sufficiently large gap and antiferromagnetic insulators with low-frequency magnetic excitations. For example, superconducting resonators which are fabricated of  Nb alloys could allow to reduce considerably their critical length, in comparison with Al wires discussed in Ref.\cite{Mooij}. Probably, easy plane antiferromagnetic insulators having, in addition, the low-energy in-plane anisotropy might be good candidates. Moreover, the microwave cavity may  differ from MS plasmon resonator (some of such systems are reviewed in \cite{cavities}). The bilayer film, which provides the coupling between the supercurrent and magnons, might be integrated in any resonator. One more way is to modify the set-up in Fig.1 by incorporating there  LC elements and Josephson junctions, or flux qubits. With nonlinear circuit elements it becomes possible to tune the frequency of the resonator. There are many possibilities for a practical realization of a proposed here mechanism. On the other hand, the goal of this work was to evaluate the strength of  coupling between cavity modes and magnons within the  proposed mechanism. Therefore, a simplest model was employed.


\appendix

\section{Calculation of a coupling between magnons and the phase of the order parameter}

In this section the parameters $\pi_{s\theta}$ and $\pi_{ss}$ will be calculated for a clean superconductor.  Before expansion of $\ln G^{-1}$ in Eq.(\ref{action}) over small perturbations it is convenient to modify $G$ with the help of the unitary transformation $G^{-1}\rightarrow U^+G^{-1}U$, where $U=\exp(i\theta\tau_3/2)$. Further, let us represent $G^{-1}$ in the form $G_0^{-1}+\mathcal{V}$, where the unperturbed function $G_0^{-1}$ is given by
\begin{equation}\label{G0A}
G_0^{-1}=\frac{\partial}{\partial\tau}+\tau_3\frac{\hat{k}^2}{2m}+\tau_3\alpha(\sigma^x \hat{k}_y-\sigma^y \hat{k}_x)-\tau_3\mu  +
\tau_1|\Delta|
\end{equation}
and the perturbation field
\begin{eqnarray}\label{V}
&&\mathcal{V}=J\bm{\sigma}\delta\mathbf{S}-i\tau_3e\left(V-\frac{1}{2e}\frac{\partial\theta}{\partial\tau}\right)+\nonumber \\
&&\frac{1}{2}\{\hat{\mathbf{k}},\mathbf{v}_s\}+\frac{1}{2}\tau_3m\mathbf{v}_s^2+\alpha m(\sigma^xv^y_{s}-\sigma^y v^x_{s})\,,
\end{eqnarray}
where
\begin{equation}\label{vs}
\mathbf{v}_{s}=\frac{1}{2m}(\bm{\nabla}\theta-\frac{2e}{c}\mathbf{A})\,.
\end{equation}
This situation has been considered in Refs.[\onlinecite{Zaikin1996,Golubev}] without the exchange and spin-orbit interactions. The latter become important in the second order expansion with respect to $\mathcal{V}$, which has the form $\mathrm{Tr} \ln G^{-1(2)}=-\frac{1}{2}\mathrm{Tr} (G_0\mathcal{V})^2$. They generate the third and fourth terms in the effective action  Eq.(\ref{Afin}). The corresponding coefficients $\pi_{s\theta}$ and $\pi_{\theta\theta}$ are given by the susceptibilities
\begin{eqnarray}\label{pi}
\pi_{s\theta}(\Omega_n,\mathbf{q})&=&\frac{T}{4}
\sum_{\omega_n, \mathbf{k}}\mathrm{Tr}[G^0_{\mathbf{k}}(\omega_n)\frac{\tilde{k}_x}{2m}G^0_{\mathbf{k+q}}(\omega_n+\Omega_n)J\sigma_y] \nonumber \\
\pi^{ij}_{\theta\theta}(\Omega_n,\mathbf{q})&=&\frac{TJ^2}{4}
\sum_{\omega_n, \mathbf{k}}\mathrm{Tr}[G^0_{\mathbf{k}}(\omega_n)\sigma^i\times \nonumber \\
&&G^0_{\mathbf{k+q}}(\omega_n+\Omega_n)\sigma^j] \,,
\end{eqnarray}
where $\tilde{k}_x=k_x-\alpha\sigma_y$, $\tilde{k}_y=k_y+\alpha\sigma_x$ and $G^0_{\mathbf{k}}(\omega_n)$ is the Matsubara Green function. For a clean superconductor it is given by
\begin{eqnarray}\label{G02A}
&&G^0_{\mathbf{k}}(\omega_n) =  - \frac{(i\omega + \xi_{\mathbf{k}}^ + \tau _3+\tau_1|\Delta|)}{\omega_n ^2 + (\xi_{\mathbf{k}}^ + )^2+\Delta^2}\frac{(1 + \bf{n}\bm{\sigma})}{2} -\nonumber \\&&
\frac{(i\omega + \xi_{\mathbf{k}}^ - \tau _3+\tau_1|\Delta|)}{\omega_n ^2 + (\xi_{\mathbf{k}}^ - )^2+\Delta^2}\frac{(1 - \bf{n}\bm{\sigma})}{2}
\end{eqnarray}
where $\xi_{\mathbf{k}}^{\pm}=k^2/2m \pm \alpha k-\mu$, $n_x=k_y/k$ and $n_y=-k_x/k$. The projection operators $(1 \pm \bf{n}\bm{\sigma})/2$ in Eq.(\ref{G02A}) split the Green function into two parts which correspond to different spin helicities of electron bands. We will assume that SOC strength is large, so that $k_F\alpha \gg \Delta$ in Eq.(\ref{G0A}). Therefore, in the leading approximation with respect to $\Delta/\alpha k_F$  one may keep in Eqs.(\ref{pi}) only terms with equal helicities of two Green functions entering in these expressions. At the same time, it is assumed that $\alpha k_F \ll \mu$. Also, the magnon frequency is much less than $\Delta$ and characteristic wave numbers $q$ of magnon and plasma modes  are much less than the superconductor's coherence length. Therefore, we set $(\Omega, q) \rightarrow 0$ in Eqs.(\ref{pi}). The expressions for the susceptibilities $\pi_{s\theta}$ and $\pi_{ss}$, which are presented in the main text, were calculated at the low temperature $T\ll  \Delta$.

The above calculations are valid for a clean system. The susceptibilities for a dirty superconductor at large SOC, such that $ \alpha k_F \gg \tau_e^{-1} \gg \Delta$, where $\tau_e$ is the elastic scattering time, can be found in \cite{Malsh vortex}.


\begin{thebibliography}{99}
\bibitem{capacitance}
For instance, if the grounded metal film is placed near the superconducting film at the distance less than $d \ll W<L$, where $L$ and $W$ are the length and the width of the film, this capacitance is given by $C=\epsilon W/4\pi d$. $\epsilon$ is the dielectric constant of a material between films.
\end{thebibliography}
\end{document}